\documentclass[journal,comsoc]{IEEEtran}
\usepackage[T1]{fontenc}% optional T1 font encoding
\usepackage{amsmath}
\usepackage[cmintegrals]{newtxmath}
\usepackage{graphicx}
\usepackage{acro}
\usepackage{todonotes}
\usepackage{url}
\IEEEoverridecommandlockouts
\usepackage{cite}
\usepackage{amsmath,amssymb,amsfonts}
\usepackage{algorithmic}
\usepackage{graphicx}
\usepackage{textcomp}
\usepackage{xcolor}
\def\BibTeX{{\rm B\kern-.05em{\sc i\kern-.025em b}\kern-.08em
    T\kern-.1667em\lower.7ex\hbox{E}\kern-.125emX}}

\usepackage{xspace}
\newcommand{\inginious}{INGInious\xspace}
\begin{document}

\title{Scaling Networking Education with Open Educational Resources
}

\author{
\IEEEauthorblockN{O. Bonaventure, Q. De Coninck, F. Duchene, M. Jadin, F. Michel, M. Piraux, C. Poncin}\\
  \IEEEauthorblockA{\textit{ICTEAM, UCLouvain} \\
    \textit{Louvain-la-Neuve, Belgium}}
  \and

    \IEEEauthorblockN{O. Tilmans}\\
  \IEEEauthorblockA{\textit{Nokia Bell Labs} \\
    \textit{Antwerp, Belgium}}

  %   email address}

  % \IEEEauthorblockN{1\textsuperscript{st} Given Name Surname}
  %   \textit{name of organization (of Aff.)}\\
  %   City, Country \\
  %   email address}
  % \and
  %   \textit{name of organization (of Aff.)}\\
  %   City, Country \\
  %   email address}
  % \and
  %   \textit{name of organization (of Aff.)}\\
  %   City, Country \\
  %   email address}
  % \and
  %   \textit{name of organization (of Aff.)}\\
  %   City, Country \\
  %   email address}
  % \and
  %   \textit{name of organization (of Aff.)}\\
  %   City, Country \\
  %   email address}
  % \and
  %   \textit{name of organization (of Aff.)}\\
  %   City, Country \\
  %   email address}
}

\maketitle

\begin{abstract}
To reflect the key role played in our society by the network technologies, the
networking courses have moved to Bachelor degrees where they are taught to
large classes. We report our experience in developing an open-source ebook that
targets those introductory networking courses and a series of open educational
resources that complement the ebook.
% The network technologies play a key role in our society and the networking courses have moved to Bachelor degrees where they are taught to large classes. We report our experience in developing an open-source ebook that is targeted at those introductory networking courses and a series of open educational resources that complement the ebook.
\end{abstract}


\section{Introduction}

% networking textbooks and the evolution of networking education over the last twenty years

% discussion of the top-down and bottom-up approaches with their advantages and drawbacks and pointers to the different books and the different approaches

In less than half a century, computer networks have revolutionized our society. The first packets were exchanged between ARPANet nodes almost fifty years ago. A few years later, the Local Area Networks era began with the early Ethernet networks at Xerox PARC. Thirty years ago, Tim Berners-Lee and his colleagues introduced the world-wide web which enabled a larger population to access the Internet. This boom started with dial-up modems in the late 1990s. Given the huge demand, network operators deployed multiple broadband access technologies, such as xDSL or cable modems. A few years later, thanks to the new mobile networks and the smartphones, it became possible for almost anyone to access the Internet at %almost
any time, % and almost
everywhere. The Internet is now part of our life and even the Human Rights Council of the United Nations has recognized the importance of the Internet.

Internet technology has evolved in parallel with its world-wide deployment. The first networking courses were graduate level courses that were aimed at training the researchers who were developing the new networks and protocols. Bertsekas and Gallager's textbook~\cite{bertsekas1992data} is an example of such graduate-level courses. The eighties and late nineties saw lots of efforts in standardizing computer networks. The OSI reference model~\cite{day1983osi} became a standard way to organize networking courses, and was adopted by several books~\cite{halsall1985introduction,tanenbaum2013computer}.

During the nineties, a growing number of universities included networking courses in their curriculum and new ways of teaching computer networks emerged. Many textbooks adopted a bottom-up approach that starts from the physical layer and moves progressively up to the application layer. This constructive approach was
popularized by Andrew Tanenbaum~\cite{tanenbaum2013computer} and has been used by generations of students. It worked well when the students were not exposed to real computer networks. Since the late 1990s, most students have been exposed to the Internet. For such students, and even more for today's ones, a top-down approach is much more motivating. This approach has been popularized by Jim Kurose and Keith Ross~\cite{kurose2017computer}. Computer Networking concepts are explained by starting from the application layer by leveraging the practical knowledge that the students already have. Then, the course can dissect the lower layers and reveal the main principles and protocols used on the Internet.
Peterson and Davie focused on the system aspects of computer networks~\cite{peterson2011computer}.

During the last ten years, we have written and expanded an open-source networking textbook~\cite{cnp3} that has been adopted by various universities around the world.
The first edition of our open-source networking textbook used the top-down approach%~\cite{cnp3}.
After a few years, feedback from the students indicated that they had difficulties with this approach. When a layer is explained using this approach, students need to learn at the same time the basic principles and algorithms and details of the deployed protocols. Many of them had difficulties to separate the key principles from the details of the protocols. We revised the course organization and the second edition of the textbook adopted a hybrid approach.

The first half of the course introduces all the key principles of computer networking using a bottom-up approach. It begins with a brief description of the physical layer. It then describes the principles of reliable delivery protocols using an abstract protocol that resides in the datalink layer. We then introduce the different organizations of the network layer and the separation between the control plane and the data plane. The students learn the principles behind the classical link-state and distance vector routing algorithms without looking at protocol specific issues. The course then explores the transport layer, showing why transport protocols are more complex than datalink layer ones and explaining the need for congestion control. Finally, the key principles for network security conclude the first half of the course.

The second part of the course adopts a top-down approach that uses the key Internet protocols to illustrate how the principles described earlier are used in the global Internet. The students learn the practical aspects through a combination of projects, labs and web-based tools.

In this paper, we report on our experience in developing various teaching materials to supplement an introductory networking course. This paper is organized as follows. We first describe in Section~\ref{section:hum} how to encourage student participation in large classes. Then, in Section~\ref{section:online}, we report our experience with a set of online open education resources that enable students to better understand various networking concepts. In Section~\ref{section:projects}, we present two projects that enable students to better understand how networking protocols are implemented and deployed. To quantify the impact of these pedagogical innovations, we sent a survey students who followed our introductory course. We received 128 replies, mainly from this year's student's, but also from students who followed the course during the previous two years. We provide the key findings of this survey throughout the text. Finally, we conclude with a discussion of some lessons we learned while developing those open educational resources.

\section{Humming in large classes}\label{section:hum}

One of the challenges of teaching to large classes is to encourage the
students to interact during the class. Today's professors do not
read their course to passive students anymore. Most professors try to initiate interaction
with the students by asking simple questions. However, many students are
shy, and they usually refrain from raising their hands to answer questions.
Unfortunately, this decreases their engagement and after some time, they even
stop listening to the teacher's questions. Several colleagues have experimented
with various types of voting systems to overcome this issue, from stand-alone systems to applications running on the student's smartphones. The former require to distribute the
voting system before the class, which consumes time. The latter does not
suffer from this problem, but once students have opened their smartphone to answer a question, they spend a few minutes to interact with other applications and lose focus.
This unfortunately reduces their engagement.

The Internet Engineering Task Force (IETF) faced a similar problem to handle votes inside working
groups when standardizing networking protocols. An IETF working group is responsible for a given protocol, and participants meet roughly three times per year. During these meetings, engineers and protocol experts discuss the new protocol specifications being developed. From time to time, working group chairs need to evaluate whether the working group agrees with one proposal. This can be done through a show of hands, but this enables every working group participant to determine who favors a given proposal. This is not always desirable. The IETF relies on \emph{humming} to solve this problem \cite{rfc4677}. Instead of asking the participants to raise their hands, the working group chairman asks the participants to hum for option A or B. If the noise level is high for A and low for B, option A is accepted. If the same noise level is heard for both options, then there is no consensus and the working group continues to discuss. Furthermore, if the noise level is low for both A and B, then this indicates that the working group participants do not have an opinion yet and the question should be asked again later.

Humming works extremely well to interact with large classes. By asking such questions every ten to twenty minutes, students remain engaged and improve their understanding of the teaching material.
Out of the 128 students who participated in a survey, 80\% always or frequently attended the lecture. Among them, 77\% totally agree or agree that humming helped them to get involved and 75\% that it drew their attention to the key concepts.

\section{Online Exercises enable students to learn from their errors}\label{section:online}

%Our open-source ebook \cite{cnp3} is available in different formats: pdf, HTML and epub. The first version was structured as a regular book that was distributed online. As indicated by various studies, students learn better from printed books where they can take notes and underline important parts than from pure online sources.

% https://www.businessinsider.com/students-learning-education-print-textbooks-screens-study-2017-10?r=US&IR=T

Over the years, we added online features to our networking ebook~\cite{cnp3}.
The first modification was to add hyperlinks to all RFCs and articles cited in the bibliography. This was a simple change, but shortly after we were positively surprised that a growing fraction of students asked questions about some of these references and cited them correctly in their project reports.
55\% of the surveyed students confirm that they follow (always or frequently) links to find additional information.

%This did not happen in the past with regular books. With an ebook, a student can easily follow a link to learn more or get additional information. In a printed book, he/she has to go to a library or start a search on the web.

We then integrated the ebook with the \inginious code-grading platform \cite{inginious}.
\inginious is an open-source project that enables professors to propose various types of exercises to students, and to automatically provide feedback and grade their answers.
Most \inginious exercises require the students to write code in a programming language. This code is then compiled and graded by unit tests that verify whether the
student correctly answered the question.
\inginious exercises are implemented as scripts that receive answers provided by a student and compute the desired feedback. As such, \inginious also supports
open questions (e.g., where the answers are short numerical values) and even
multiple-choice questions.

Students not only learn by reading the teaching material in the ebook or
listening to their professors or teaching assistants. They also learn by answering
questions. An important benefit of having an online ebook is that it enables the students to interact in different ways. Throughout the years, we have developed different types of exercises that enable the student to test their knowledge of the teaching material while learning it. While they are learning a new topic, students make mistakes. These mistakes are part of their learning process. Students can benefit from these mistakes provided that they receive fast and accurate feedback. Our online ebook includes different types of exercises, from simple multiple choice questions to more complex programming assignments that all provide immediate and precise feedback to the students.

\subsection{Multiple choice questions}

\begin{figure}
  \center%
  \includegraphics[width=0.45\textwidth]{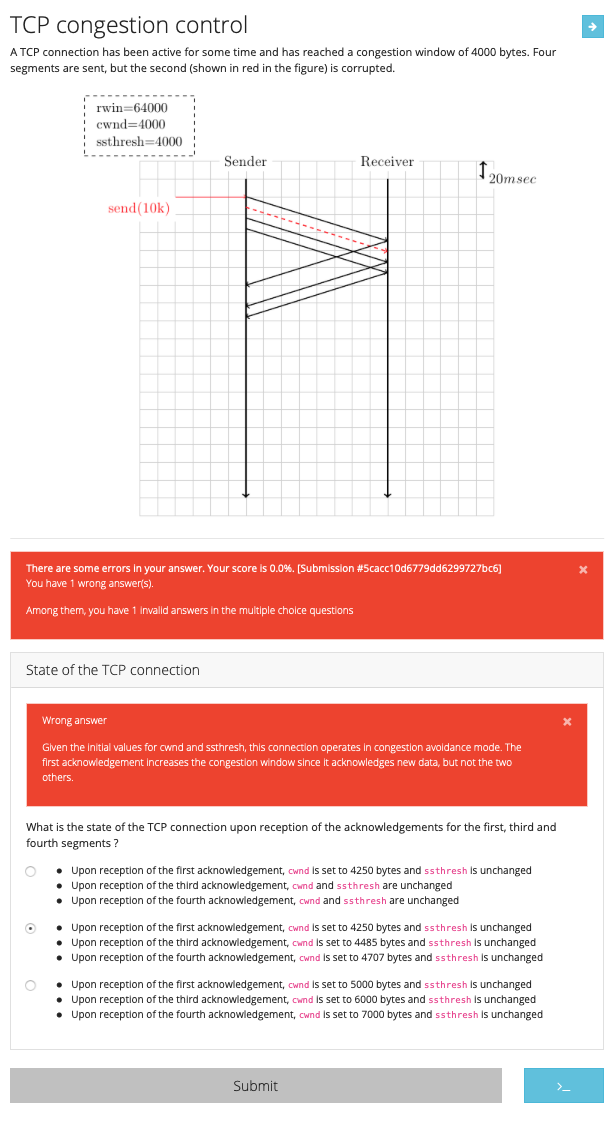}
  %TODO: (MP): This one is hard to read, maybe drop one of the choices and cut/wrap the long sentences}
  \caption{Multiple choice questions provide immediate feedback to students and clarify common misconceptions.}
  \label{fig:mcq}
\end{figure}

Our ebook includes dozens of interactive multiple choice questions that cover a wide range of topics. In order to encourage the students to answer these questions several times (e.g., during their first read, and later when they prepare the exam) each question contains two sets of positive and negative answers. When displayed,
each questions is then dynamically constructed by a JavaScript module that randomly
selects $one$ correct answer and $n$ invalid ones.
To provide feedback, each possible answer has an associated comment. As the student
selected his/her answer, the corresponding comment is then shown.
For invalid answers, the comment thus provides the explanation that enables the
student to correct his/her misunderstanding. One example of such questions is
shown in Figure~\ref{fig:mcq}.
58\% of our students report  they answered (always or frequently) the multiple choice questions directly while reading the ebook, and 88\% during the semester. 73\% reused them while preparing for the exam.

\subsection{Short answers}

\begin{figure}
  \center
  \includegraphics[width=0.45\textwidth]{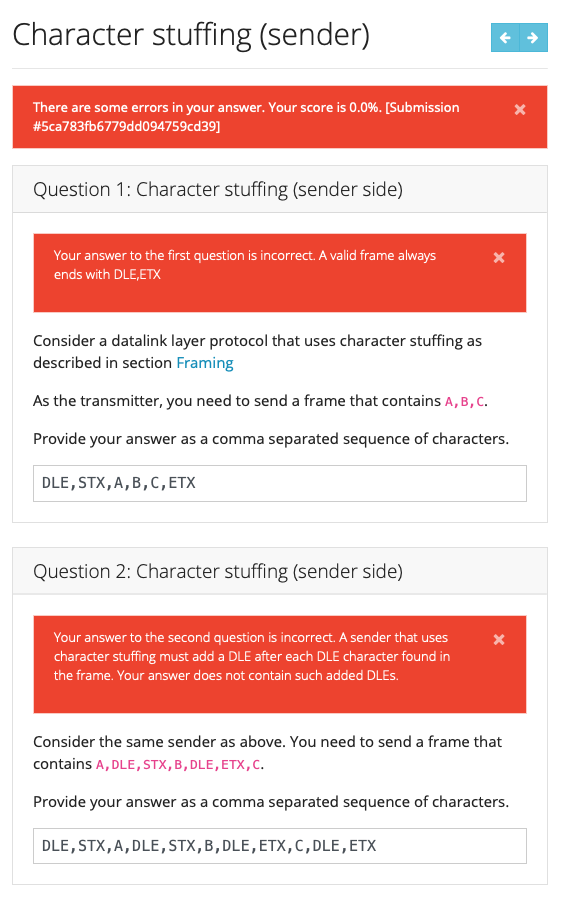}
  \caption{Questions with a short answer provide immediate feedback to the students.}
  \label{fig:short}
\end{figure}

The multiple choice questions do not allow testing all the topics described in a basic networking course. For some concepts, it is interesting to ask questions that require a short answer, typically a number or a few words. Many simple questions can be framed this way and it is more challenging for a student to answer such questions than to guess one answer among $n$ choices. We use such open \inginious questions to cover topics where the students can provide a short answer. One example is shown in Figure~\ref{fig:short} where the student needs to provide the frame that will be sent when character stuffing is used. \inginious provides detailed feedback that enables the students to learn from their mistakes.

\subsection{Socket programming}

Most networking courses include an introduction to socket programming. This low-level API is the classical way to develop networked applications that interact with the transport layer, mainly TCP and UDP. Several textbooks \cite{stevens2004unix,beej} describe these interactions in details. Despite the qualities of these textbooks, many students have difficulties in understanding several of the key concepts that underlie the development of networked applications. Students have difficulties in understanding the difference between big-endian and little-endian. Many of them forget to check the return values of system calls or cannot correctly parse data received from the network.

As an example, let us illustrate one of these questions that verifies whether a student correctly uses the socket layer to interact with UDP. The students need to implement a client that sends a vector of integer in a UDP datagram to a server and receives a UDP datagram with the sum of these integers. Both the client and the server are implemented in one function whose correct operation is verified by \inginious. These questions are verified with a script that compiles the C code and then runs a series of unit tests. These unit tests validate different aspects of the code that the students have to write. Some of these tests include simple assertions that verify that the correct value is returned for a given set of parameters. Others are more subtle. For example, to verify that the students correctly check the return values of the memory allocation functions, our tests use library wrappers to intercept these functions and force them to return errors in some tests. We do the same for system calls such as \texttt{send} and \texttt{recv}. Our tests also verify that the students use the network byte ordering when sending and receiving integers.

We developed a similar exercise that uses TCP instead of UDP. In this case, the client opens a connection to the server, sends its vector of integers and waits for the answer. Our tests verify that the client and server codes correctly use the socket interface. Some of our tests cover corner cases that students often ignore at a first glance. For this exercise, many students expect that when a server issues a \texttt{recv} system call, it will also return the entire vector of integers. This is what they usually observe from a small test in the LAN on their computer. In reality, the \texttt{recv} system call returns the number of bytes that have been received in-sequence. If the vector was sent in multiple packets on a WAN, only a fraction of them could be available when the server calls \texttt{recv}. Some of our tests simulate this behavior to %remind the
encourage the students to immediately write code that handles all %error
possible cases from day one.

\subsection{Learning Internet protocols}\label{section:wireshark}

One of the objectives of many networking courses is to enable the students to understand how network protocols operate. Most textbooks include a textual description of the protocol that includes the packet format, a finite state machine and some examples. This textual description is fine for the assiduous students, but many protocols rely on conventions that are important and need to be well understood by the students. While these conventions can be explained in text or with some pseudo-code, students better learn about these conventions by observing and interacting with the protocols. For many students, using a packet dissector like \texttt{Wireshark} or \texttt{tcpdump} makes the network protocols become real, and they learn by observing the succession of packets that are exchanged inside a network.

\begin{figure}
  \center
  \includegraphics[width=0.45\textwidth]{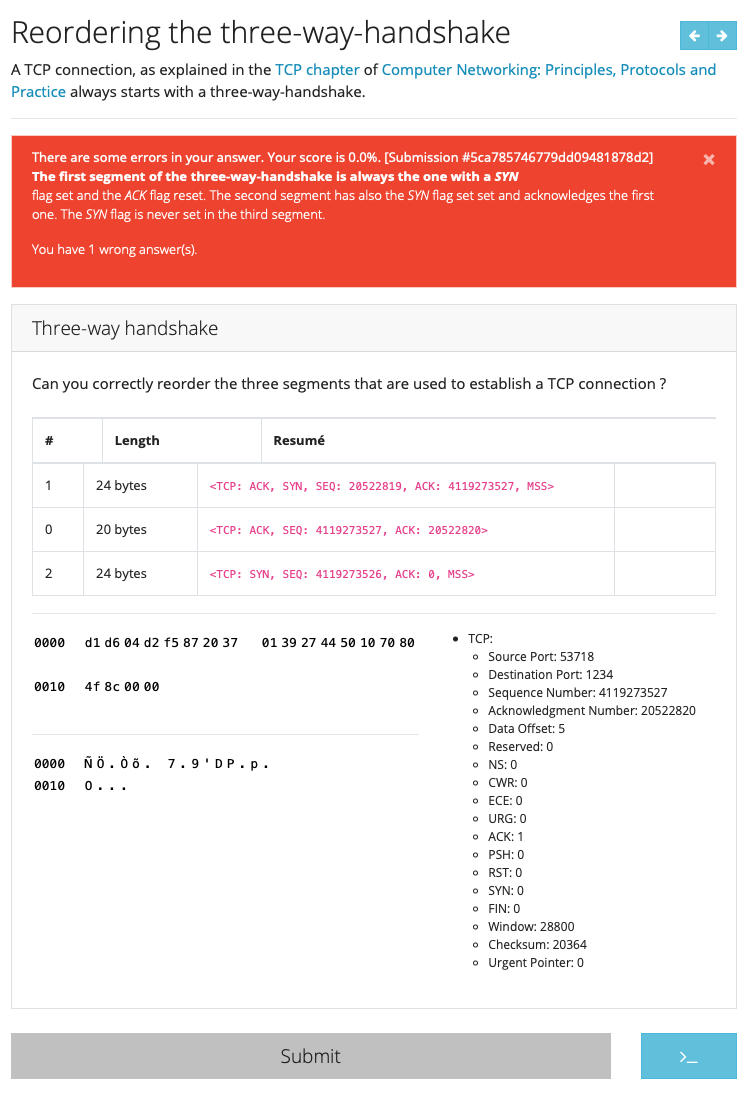}
  \caption{Students learn the TCP three-way handshake by reordering packets.}
  \label{fig:tcp-handshake}
\end{figure}

Each of the fields that are included in packet headers have a specific role. Some of them have different roles that depend on the values of specific flags. To reinforce their understanding, our ebook includes a series of \inginious exercises that require the students to predict the value of specific fields of packets in a given exchange. To develop such exercise, a professor simply needs to collect packets in a real or emulated network and select the interesting ones. We modified \inginious to include a simplified packet dissector that presents the packet trace similarly to \texttt{Wireshark}. The student can look at the content of each packet and observe the values of the different fields. However, some fields of the packet header are masked and the student needs to predict their values based on the context provided by the other packets in the trace. A simple example is provided in Figure~\ref{fig:tcp-handshake} where the students have to reorder the first three packets of TCP's three-way handshake.
Indeed, 66\% of the students who used this tool found it useful.

%This Figure
Figure\ref{fig:tcp-handshake} shows the three packets that compose a TCP handshake in random order. Packet \#\texttt{2} is the \texttt{SYN} packet sent by the client. Packet \#\texttt{1} is the \texttt{SYN+ACK} that was returned by the server and packet \#\texttt{0} is the third packet of the handshake. In this exercise, \inginious randomises the packet trace and the students have to reorder them. In this example, the student did not reorder the packets correctly and \inginious returned feedback.

%the Acknowledgment Number of the SYN+ACK packet has been masked and the student needs to compute its value based on the information found in the other packets. In this example, the server must return as Acknowledgment Number the next byte that it expects to receive in sequence after having received the \texttt{SYN} flag that consumed Sequence Number $4207062859$ in the initial packet.

%Such exercises can be added easily to the ebook or designed by teaching assistants. The teacher simply needs to provide a short trace, typically a few packets and list the fields that must be masked by the web interface in each specific packet (if any). We have developed such packet exercises that cover various aspects of TCP, IPv6 or ICMP6.

\subsection{Virtual Labs}

The \inginious exercises described in the previous sections enable the students to learn the basic
principles of the Internet protocols. However, this is not sufficient to let them grasp how the protocols behave in real networks. Nothing replaces experiments with real protocol implementations. Many universities and enterprises
use labs equipped with routers, switches servers and many cables. However, given the growth in the number of
Computer Science students, it became impossible for us to manage and assign such physical labs for two hundred students. %With
%such large numbers, it became impossible to assign students to physical labs.
As other universities \cite{netkit},
we replaced the physical labs with virtual labs that enable the students to emulate simple networks. In particular,
Mininet \cite{mininet} leverages the Linux namespaces to efficiently emulate several hosts, switches and routers on
a single Linux host. Thanks to Mininet, the students can conduct simple tests with switches and hosts.

When the students learn the interactions between the TCP congestion control and the retransmission techniques, we
regularly ask them to predict how and when packets will be sent in a given scenario. For example, consider a
TCP connection between two hosts that are connected over a network with a 20 msec round-trip-time. The client
retrieves 8 packets from the server and the sixth and eight are lost. The students predict the packets that will be sent
by the server according to the New Reno congestion control scheme and then verify their assumption by comparing their
prediction with measurements in a Mininet lab using \texttt{tc} and eBPF to emulate a deterministic packet dropper.

Over the years, we have extended Mininet to simplify the configuration of routing protocols. Many students have
difficulties to interact with the routing protocol implementations that are supported by Linux. \texttt{IPMininet} \cite{ipmininet} is a set
of Python classes that extend Mininet and provide a simple API that allows professors to define network topologies
where different protocols are used. These are exposed to the students who can then capture packets or change configuration parameters. Some of these virtual
labs contain invalid forwarding tables
or some incorrect routing configurations that the students are tasked to debug.

%\cite{mininet}, \cite{netkit}

% ipmininet

% BGP lab

% tcp tests

\section{Student projects}\label{section:projects}

Students learn a lot through projects. These projects enable them to be creative and expand their understanding of the course topics. The main difficulty in creating student projects is to find the right balance between the time that the students spent on the project and the skills that they learn. Over the years, we have obtained very good results with two different projects. The first project is organized during the first half of the course. The students implement a simple transport protocol using the socket layer. During the second project, they analyze the architecture and performance of a popular web server.

\subsection{Implementing a simple transport protocol}

Our first project is proposed to pairs of students. Pairs of students are natural for the implementation of a client-server protocol, as one student can focus on the client side while the other focuses on the server side. During the project, the students need to develop a simple transport protocol that uses window-based flow control, acknowledgments, and different retransmissions techniques. We vary some details of the protocol every year to prevent students from simply reusing code written by their predecessors.

The project is organized in four phases and the students benefit from each of them. The first phase is the bootstrap phase. During this phase, the students learn the basic principles of the reliable protocols and write small functions in C that are tested by \inginious. At the end of this phase, the students should have learned the basics of socket programming.
61\% of our students totally agree or agree that these \inginious exercises were sufficient to allow them to start the project.
They then develop their first implementation of the simple transport protocol. This protocol runs on top of UDP for practical reasons. We provide them with a Wireshark dissector to analyze packets and a simple link simulator that introduces delays, reordering and losses on a given UDP flow. At the end of this phase, each pair of students develops a prototype implementation of the simple transport protocol. Many professors would grade the project at this stage. Our experience is that by doing so, the students would miss many lessons that experienced network engineers learn by interacting with others. When new protocols are developed, the first implementers of these protocols often organize interoperability events to validate their respective implementations. Such events are a unique learning opportunity for the students as well. We organize our interoperability tests during a week and each pair of students is required to test their implementation with at least two other implementations and report the results. Many of these tests enable the students to detect subtle bugs in their implementations and fine-tune them.
The interoperability tests are stressful for the students (71\% because it is a strict deadline) but mainly helpful: 74\% have discovered errors in their implementation, 72\% have identified ideas for improvements by observing the behavior of other implementations and 82\% modified their implementation based on these tests.

The interoperability test verifies that different implementations can interact. If the test succeeds, it is likely that the students have correctly implemented the protocol. However, this test does not tell anything about the architecture and the quality of the student's code. We use the fourth phase of the project to cover this aspect. We organize a peer-review phase during which each student evaluates two implementations from other groups of students. This peer-review is organized through a local instance of the \texttt{hotcrp} conference reviewing website. Each student has to answer a set of questions on the quality of the code, its structure and correctness. We intentionally do not ask the students to grade the project from other students in contrast with other colleagues who use peer-reviews to grade student projects. We use the peer reviews to achieve different pedagogical objectives. First, the peer reviews encourage the students to read code written by other students. Code reviews are an important part of the work of many computer scientists and it is important to train the students to these activities. We insist on constructive comments that can be used by the students who receive them to improve the quality of their project. Second, these peer reviews provide qualitative feedback to the students. Given the size of our classes, we cannot anymore provide personalized feedback to each student and have to rely on automated tests when grading student projects. Third, we allow students to update their project based on the feedback that they receive. This improves the quality of their projects and thus their grades. The reviews returned by the students are graded and contribute to a small fraction of the course grade.
This phase of peer-review pushes students to be diligent: 65\% of students want to get a positive report from their peers. According to students, peer-reviewing is useful. 74\% of our students have identified classical errors that they would avoid in the future by reviewing the code of other students. Furthermore, 73\% have discovered good practices that they plan to apply to future projects.

We have tested two strategies to allocate the peer reviews to the students. Our first strategy was to randomly allocate two reviews to each student. This strategy is simple to implement and works for most students. Its main advantage is that each project is reviewed by the same number of students. However, it leads to problems with very good or very bad student projects. For such projects, it is very difficult for an average student to provide constructive feedback. We now randomly allocate five projects to each student and then let him/her select the two of them that he/she will review. This avoids difficulties with projects that are too weak or too strong, but does not guarantee that all projects receive feedback from other students. As we have no guarantee that all students write the required reviews anyway, this problem is also present with the initial allocation strategy.

%hotcrp

%reviews puis resoumission

%evaluation

\subsection{Observing deployed protocols}

During the second half of the networking course, we organize a six weeks long project that encourages the students to improve their understanding of key Internet protocols. This project starts shortly after the presentation of the application layer and progresses in parallel with the course. For this project, each student selects one website and provides a short but precise four-pages two column paper describing how various Internet protocols have been tuned on this website. Initially, we encouraged each student to select one website to analyze. This offered a strong motivation for the students as they studied a website that they used frequently. However, grading those projects was difficult since all students observed different optimizations. We now ask the students to vote for their most popular websites and select a subset of them so that each website is analyzed by about a dozen of students. This provides a good diversity in what students can observe and simplifies the grading, which is important with classes of 200 students and more.
73\% of our students confirm that observing protocols on real websites was motivating and pushed them to surpass themselves.

Every week, we encourage the students to focus their analysis on one particular aspect of the studied website. We provide guidelines and suggests tools that the students could use at the beginning of each week. We start from the simpler protocols and then expand to more complex protocols or those that require special tools. Before this project, the students have been exposed to packet dissectors such as \texttt{Wireshark} and the exercises described in Section~\ref{section:wireshark}. We start usually start by analyzing the configuration of the DNS of studied website. With simple tools such as \texttt{dig} or \texttt{nslookup}, the students can determine the number of nameservers, the types of DNS records, the support for IPv4 and/or IPv6. Using \texttt{traceroute}, they try to determine where the website is hosted. We also provide to the students RIPE Atlas measurements from probes located on different continents to the studied websites. This gives some hints on how those websites rely on Content Distribution Networks. %During the recent years, the students have observed the growing importance of IPv6 and DNSSec using these measurements.

The second week is devoted to HTTP. We leverage the developer extensions of the Chrome, Firefox, and Safari browsers and ask the students to analyze first the number of servers that are contacted to visualize a given web page. Students are usually puzzled when they first observe that dozens of websites are usually contacted to render a single web page. They then analyze the HTTP headers to identify the non-standard ones and try to infer their meaning. Another element for the HTTP analysis are the HTTP cookies. The students are usually very surprised by the real utilization of these cookies. %During the last years, they have also started to observe the deployment of HTTP/2.

The third week is devoted to the analysis of Transport Layer Security (TLS). Ten years ago, TLS was reserved for banks and small parts of e-commerce websites. During the last years, we have observed an explosion of the number of websites that use TLS. During this week, the students use either the developer extensions of their browsers or \texttt{openssl}'s \texttt{s\_client} to analyze how websites negotiate TLS (e.g.,
version number, proposed cryptographic schemes, and specific parameters). This analysis highlights the complexity of the TLS configuration of large websites and the trade-offs that they have to do. %They have started to observe the deployment of TLS 1.3.

During the fourth week, the students observe how TCP is used by large web servers. They analyze the TCP options advertised by the server, verify whether it supports Explicit Congestion Notification, or try to infer its initial congestion window.

After this analysis, the students finalize their report. Our schedule does not enable us to organize a peer-review round for this project, but this would be very valuable. Each student report is reviewed by one professor who provides a short feedback by email and a grade. The best project reports are published on our department website and used as examples during the next year.

An important point of this analysis is that students directly observe that the Internet evolves. Five years ago, they observed websites that were using HTTP version 1.1 over IPv4. Today, they observe websites that rely on HTTP version 2.0 over TLS and IPv6 and some have already observed %the
first deployments of QUIC.

\section{The benefits of open-source}

The Internet uses a wide range of open-source software from operating systems like Linux or FreeBSD that implement the TCP/IP protocol suite to client and server applications. When we started to write the open-source \emph{Computer Networking: Principles, Protocols and Practice} ebook a decade ago, our motivation was to contribute back to the community. Over the years, it has benefited from various contributions by students and colleagues.
Students report that having access to the book sources encourages them to contribute
to its improvement (35\%) and to look at other open-source projects (37\%), which
is an unusual way to expose them to the open-source movement.

Students do not anymore simply learn networking by reading textbooks. They need a variety of activities to master this complex topic. We have developed and released various open education resources which can be used during networking classes. A key element of these resources is that they enable the students to learn from their mistakes, from the simple multiple choice questions that are integrated in the ebook, providing instantaneous feedback to the students, to the interoperability tests and the peer-reviews that encourage the students to provide constructive feedback.

Internet protocols have evolved thanks to the availability of open-source implementations that have been largely deployed or used as the starting point to develop commercial implementations.
The same applies to networking education. Open-source enables the sharing of educational resources. We encourage other colleagues to contribute to the resources that we have assembled on \url{https://www.computer-networking.info}.

% What have we learned during the last twenty years of teaching networking

% what are the next

% lab and lifelong learning

\section*{Acknowledgements}

We would like to thank all the students, teaching assistants and colleagues who have provided comments and suggestions that have encouraged us to improve our networking course and the open educational resources described in this paper. Jean-Martin Vlaeminck and Mathieu Xhonneux implemented the socket exercises.

\bibliographystyle{unsrt}
\bibliography{books}

\begin{thebibliography}{10}

\bibitem{bertsekas1992data}
Dimitri Bertsekas and Robert Gallager.
\newblock {\em Data Networks}.
\newblock Prentice-Hall Internat. Ed. Prentice-Hall International, 1992.

\bibitem{day1983osi}
John Day and Hubert Zimmermann.
\newblock The {OSI} reference model.
\newblock {\em Proceedings of the IEEE}, 71(12):1334--1340, 1983.

\bibitem{halsall1985introduction}
Fred Halsall.
\newblock {\em Introduction to data communications and computer networks}.
\newblock Addison-Wesley Longman Publishing Co., Inc., 1985.

\bibitem{tanenbaum2013computer}
Andrew Tanenbaum and David Wetherall.
\newblock {\em Computer Networks}.
\newblock Pearson. Pearson, 2013.

\bibitem{kurose2017computer}
Jim Kurose and Keith Ross.
\newblock {\em Computer Networking: A Top-Down Approach, Global Edition}.
\newblock Pearson Education Limited, 2017.

\bibitem{peterson2011computer}
Larry Peterson and Bruce Davie.
\newblock {\em Computer Networks: A Systems Approach}.
\newblock The Morgan Kaufmann Series in Networking. Elsevier Science, 2011.

\bibitem{cnp3}
Olivier Bonaventure et~al.
\newblock Computer networking : Principles, protocols and practice, 2009.
\newblock \url{https://www.computer-networking.info} - Accessed April-10-2019.

\bibitem{rfc4677}
Paul Hoffman and Susan Harris.
\newblock The {Tao} of {IETF}: A novice's guide to the {Internet Engineering
  Task Force}.
\newblock RFC4677, 2006.

\bibitem{inginious}
Guillaume Derval, Anthony Gego, Pierre Reinbold, Benjamin Frantzen, and Peter
  Van~Roy.
\newblock Automatic grading of programming exercises in a {MOOC} using the
  {INGInious} platform.
\newblock {\em European Stakeholder Summit on experiences and best practices in
  and around MOOCs (EMOOCS'15)}, pages 86--91, 2015.

\bibitem{stevens2004unix}
Richard Stevens, Bill Fenner, and Andrew Rudoff.
\newblock {\em {UNIX} network programming}, volume~1.
\newblock Addison-Wesley Professional, 2004.

\bibitem{beej}
Brian Hall.
\newblock {\em Beej's Guide to Network Programming}.
\newblock Jorgensen Publishing, 2011.

\bibitem{netkit}
Maurizio Pizzonia and Massimo Rimondini.
\newblock Netkit: network emulation for education.
\newblock {\em Software: Practice and Experience}, 46(2):133--165, 2016.

\bibitem{mininet}
Bob Lantz, Brandon Heller, and Nick McKeown.
\newblock A network in a laptop: rapid prototyping for software-defined
  networks.
\newblock In {\em Hotnets 2010}, 2010.

\bibitem{ipmininet}
Olivier Tilmans and Mathieu Jadin.
\newblock {IPMininet}.
\newblock \url{https://github.com/cnp3/ipmininet}, Accessed April-10-2019.

\end{thebibliography}

\begin{IEEEbiographynophoto}{Olivier Bonaventure}
  is Professor at UCLouvain where he leads the IP Networking Lab. He has actively contributed to the design, implementation and deployment of several Internet protocols including LISP, Multipath TCP, IPv6 Segment Routing, \ldots He currently serves as Editor for SIGCOMM's Computer Communication Review.
  %He wrote the \emph{Computer Networking: Principles, Protocols and Practice} open-source ebook that is used by various universities.
He co-founded the Tessares spinoff that deploys Hybrid Access Networks using Multipath TCP.
\end{IEEEbiographynophoto}

\begin{IEEEbiographynophoto}{Quentin De Coninck}
  received the M.Eng. degree in computer engineering from UCLouvain 2015.He is currently pursuing a Ph.D.% as a FNRS Research Fellow within the IP Networking Lab in the same institution.
His research interests mainly focus on multipath transport protocols. He first worked on an evaluation and an adaptation of Multipath TCP for mobile applications. He later proposed and implemented multipath extensions for QUIC.
\end{IEEEbiographynophoto}

\begin{IEEEbiographynophoto}{Fabien Duchene}
  received his M.Sc degree in computer science from Université Lille-I in2011. He is currently pursuing a Ph.D.% as. %a research assistant within the IP Networking Lab at UCLouvain.
 His research interests are mainly path selection in networks and multipath transport protocols. He contributed to the standardization and the Linux implementation of Multipath TCP. Lately, he has been working on IPv6 Segment Routing and eBPF.
\end{IEEEbiographynophoto}

\begin{IEEEbiographynophoto}{Mathieu Jadin}
	received the M.Eng. degree in computer engineering from UCLouvain in 2016.
He is currently pursuing a Ph.D. % within the IP Networking Lab in the same institution.
His research interests mainly lie in Software Defined Networks and IPv6 Segment Routing.
%He first worked on a Software Defined Network architecture based on IPv6 Segment Routing.
%Later, he designed and implemented a new traffic engineering algorithm tuned for networks
%using IPv6 Segment Routing.
\end{IEEEbiographynophoto}

\begin{IEEEbiographynophoto}{Fran\c cois Michel} received his M.Sc degree in Computer Science from UCLouvain in 2018.
He is now pursuing a Ph.D. % as a FNRS Research Fellow in the same institution under the supervision of Prof. Olivier Bonaventure.
His research interests include forward erasure correction techniques and their impact on the transport protocols. He recently proposed a forward erasure correction extension for QUIC.
\end{IEEEbiographynophoto}

\begin{IEEEbiographynophoto}{Maxime Piraux} received his M.Sc degree in Computer Science from UCLouvain in 2018.  He is now pursuing a Ph.D. %at the same university within the IP Networking Lab.
He studies the evolution of transport protocols implementations and works on improving their extensibility. He developed the QUIC-Tracker tool reporting the level of specification compliance of existing QUIC implementations.% QUIC-Tracker is the only known active measurement tool for QUIC.
\end{IEEEbiographynophoto}

\begin{IEEEbiographynophoto}{Chantal Poncin} obtained her engineering degree in applied mathematics at UCLouvain. She completed her training with a diploma in pedagogy. She advises Computer Science professors on the creation, coordination and evaluation of their teaching methods and tools. She also
  accompanies students throughout their university degress.
\end{IEEEbiographynophoto}

\begin{IEEEbiographynophoto}{Olivier Tilmans} obtained his Ph.D. from UCLouvain in 2019. He is now a researcher at Nokia Bell Labs in the End-to-End Network Service
Automation lab. His current research focus on software-oriented research for the
convergence of access networks, end-to-end network control and traffic management,
and fine-grained network monitoring.
\end{IEEEbiographynophoto}

\end{document}